\documentclass[twocolumn,prl,aps,floatfix,superscriptaddress]{revtex4}

\usepackage{color}
\usepackage{ulem}

\usepackage{graphicx}
\usepackage[all]{xy}

\usepackage{amsthm}

\newcommand{\be}{\begin{equation}}
\newcommand{\ee}{\end{equation}}

\begin{document}
\title{Self-Correcting Quantum Memories Beyond the Percolation Threshold}

\author{Matthew B. Hastings}
\affiliation{Microsoft Research, Station Q, CNSI Building, University of California, Santa Barbara, CA, 93106}
\affiliation{Quantum Architectures and Computation Group, Microsoft Research, Redmond, WA 98052 USA}

\author{Grant H. Watson}
\affiliation{Department of Physics and Astronomy, University of Waterloo, Ontario, N2L 3G1, Canada}

\author{Roger G. Melko}
\affiliation{Department of Physics and Astronomy, University of Waterloo, Ontario, N2L 3G1, Canada}
\affiliation{Perimeter Institute for Theoretical Physics, Waterloo, Ontario N2L 2Y5, Canada}
\begin{abstract}
We analyze several high dimensional generalizations of the toric code at nonzero temperature.  We find that in large enough dimension, there can be a distinct separation between the critical temperature $T_c$, given by thermodynamic singularities, and the percolation temperature $T_p$, given by the percolation of defects.  We argue that the regime $T_p<T<T_c$ is a range of temperatures where a self-correcting quantum memory can operate despite having percolating defects.  We present analytic arguments and numerical evidence in support of this scenario, including a mean-field treatment and Monte Carlo simulations.  Near $T_c$, simulations observe a large hysteretic behavior, which may have applications by allowing the self-correcting phase to survive in a ``superheated" regime.
\end{abstract}
\maketitle

The classical Ising model is the prototypical example of a ``self-correcting" memory.  In two or more dimensions, in the low temperature ferromagnetic phase, the system can store a single classical bit for an exponentially long time in the sign of the global magnetization.  
In contrast, many proposals to protect quantum information against noise require active error correction by an external classical control \cite{threshold}.  Other proposals like low-dimensional topological quantum memories must avoid thermally excitating anyons\cite{topo}, requiring a temperature that tends to zero as the inverse logarithm of system size.  

Remarkably, a topological memory in four dimensions ($4d$) \cite{kitaev4d} self-corrects up to a fixed nonzero temperature $T_c$, describing a transition between a high-temperature disordered phase, and a low-temperature phase capable of topologically encoding the quantum state.  
There have been extensive searches for lower-dimensional self-correcting memories \cite{bacon},
including the cubic code \cite{Haah}, though it still does not have 
a lifetime diverging arbitrarily largely with system size 
at $T>0$ \cite{Haah2}.  
The question of true self-correction in $d<4$ is still an outstanding open problem.
However, it has recently been proposed that toric ``surface'' codes can be artificially constructed \cite{Mariantoni}, where for example 
higher-dimensional superconducting quantum circuits might be engineered by building long-range connections between local circuit components in lower dimensions.  The imminent possibility of fabricating small-scale surface codes motivates us to analyze in more detail the properties of higher dimensional models.

Conventional understanding states that the loss of a self-correcting phase occurs at $T_c$ when thermally-activated defects
percolate, destroying the topological information.
Surprisingly, using a
combination of numerical simulation and gauge-invariant mean-field theory,
we find that 
the phase transition temperature $T_c$ of certain higher dimensional toric codes is {\it not} the same as the temperature $T_p$ for percolating defects, and that error correction can be possible even when defects percolate.  In fact, we present analytic arguments that the ratio $T_c/T_p$ can diverge as $d$ becomes large, allowing error correction over an unexpectedly large range of temperatures.  Further, Monte Carlo simulations find a large region of hysteresis, where the low-$T$ phase is metastable well above $T_c$.  This raises the possibility of combining self-correction with a small amount of additional active correction to maintain the topological phase in a ``superheated" regime.  

\begin{figure}
\begin{center}
\scalebox{1}{\includegraphics[width=3.2in]{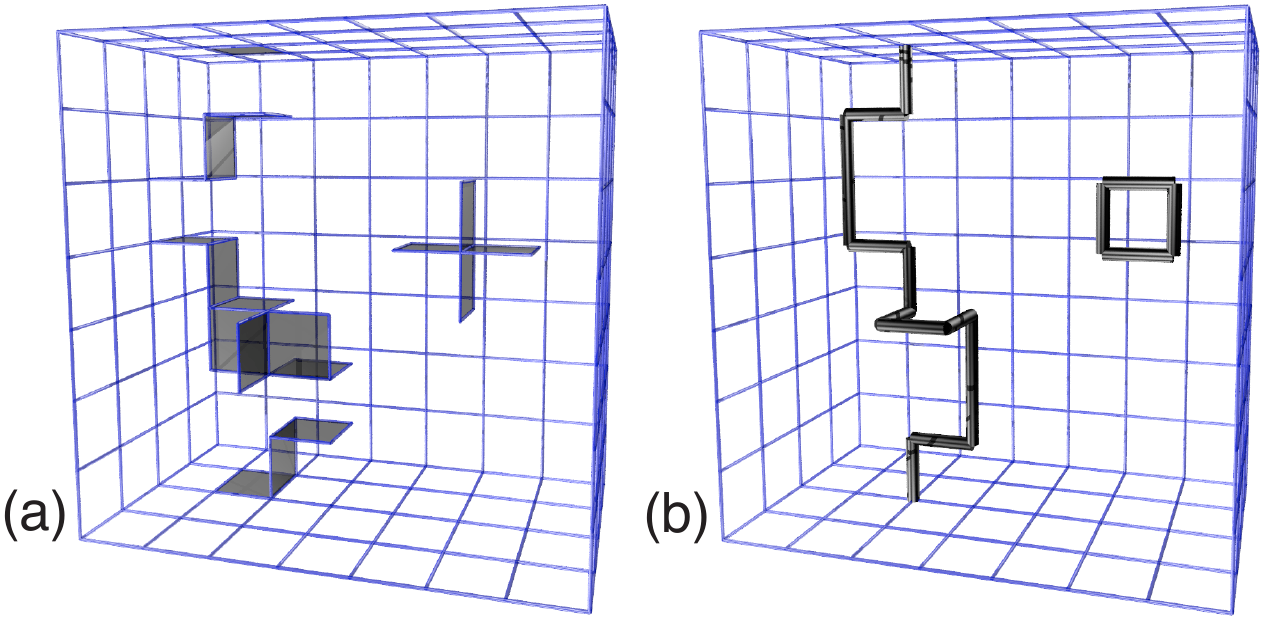}}
\end{center}
\caption{ Defect clusters in the $3d$ code. 
(a) A set of percolating (left) and non-percolating (right) clusters of defect plaquettes in $H_A$, defined as being 2-cells that have an odd number of Ising variables $S^z_i$. (b) The defect surfaces in the dual picture.}
   \label{fig:cube}
 \end{figure}

{\it $(p,q)$ Toric Codes---}
We begin by briefly describing the formalism for generalizations of the toric code to a
hypercubic lattice in $d$-dimensions.
We refer to the different codes as $(p,q)$ codes, with $p+q$ equal to the spatial dimension $d$.
In this notation, the original toric code is a $(1,1)$ code while the four dimensional self-correcting code is a $(2,2)$ code.
Following the usage in topology, we refer to the vertices of the lattices as $0$-cells.  The edges of the lattice are $1$-cells, the plaquettes are $2$-cells, and so on, up to $d$-cells.  We use $N_k$ to denote the number of $k$-cells.  Note that $N_k=N_{d-k}$; this equality is related to a duality between $(p,q)$ codes and $(q,p)$ codes.

In a $(p,q)$ code, there is one spin-$1/2$ (or qubit) on each $p$-cell.  The Hamiltonian is
$H=J_A H_A+J_B H_B,
$ %\ee
with $J_A,J_B$ being positive scalars and
\be
\label{Asum}
H_A=-\sum_{c_{p+1}} \prod_{i \in c_{p+1}} S^z_i, \hspace{5mm}
H_B= - \sum_{c_{p-1}} \prod_{i \ni c_{p-1}} S^x_i.
\ee
In $H_A$, the sum is over $p+1$-cells, denoted $c_{p+1}$, and the product is over $p$-cells, labeled by $i$, with $i \in c_{p+1}$ meaning the $i$ is attached to $c_{p+1}$.
The notation in $H_B$ is similar, with the sum being over $p-1$-cells, $c_{p-1}$.

The partition function $Z(\beta)={\rm tr}(\exp(-\beta H))$ decomposes exactly as
\be
Z(\beta)=Z_A(\beta) Z_B(\beta) 2^{-N_p},
\ee
where $Z_A={\rm tr}(\exp(-\beta H_A))$ and $Z_B={\rm tr}(\exp(-\beta H_B))$.
Hence, we can compute $Z(\beta)$ by calculating $Z_A$ and $Z_B$ separately, e.g. using classical Monte Carlo.

In this paper, we focus on $(1,d-1)$ codes for simplicity.  Strictly speaking these codes cannot be self-correcting quantum memories: they can self-correct against spin flip errors but not against dephasing errors ($S^z$ errors).  A $(2,d-2)$ code can self-correct against both types of errors for $d\geq 4$ \cite{adjust}.  The self-dual code is the $(d/2,d/2)$ code -- we will address that in a future work.
In this paper, we
only simulate $H_A$, where the spins are on $1$-cells (edges) and the interactions are on $2$-cells (plaquettes);  i.e.~we study self-correction against spin flip errors only.

The $(0,d)$ code is the Ising model, with no ability to correct against dephasing errors, being only useful as a classical memory.  Here, the phenomenon that the percolation and phase transition temperatures are distinct is well established.  It has been proven that $T_p \sim d/\log(d)$ for large $d$ \cite{PercIsing} while $T_c \sim d$ in agreement with mean-field theory \cite{MFTIsing}, and so for sufficiently large $d$, $T_p<T_c$.  Further, numerical simulations \cite{3dIsing} show that this occurs already in $3d$.
This difference between $T_p$ and $T_c$ means that the Peierls argument \cite{Peierls}, which relies on convergence of a low temperature expansion, cannot correctly predict the $T_c$ for the Ising model.

{\it Defects, Percolation Temperature, and Relation to Error Correction---}
In the Ising model, the low temperature expansion sums over domain walls, called Peierls contours, between up and down spins, with a weight dependent on the area of the contour.  It is easier to understand these contours if we go to a {\it dual} version of the Ising model, a $(d,0)$ code.  In this case, the spins are on the $d$-cells, and the interactions are on the $d-1$-cells between a pair of $d$-cells.  A spin configuration is given by assigning $+1$ or $-1$ to each $d$-cell and the set of interactions which are unsatisfied is then given by taking the boundary of this configuration (here, the boundary can be understood intuitively as the $d-1$-cells that connect two $d$-cells with opposite signs of spin, but the general definition is to use a boundary operator on a chain complex).  Since the set of unsatisfied interactions is a boundary, and the boundary of a boundary vanishes, the set of unsatisfied interactions indeed give closed surfaces.  For these surfaces made of $d-1$-cells, we regard two $d-1$-cells as being neighbors if they both attach to the same $d-2$-cell.
If we return to the $(0,d)$ code, then the defects are configurations of $1$-cells instead, and two $1$-cells are neighbors if they both attach to the same $2$-cell.

For a $(1,d-1)$ code, the defect surfaces are sets of $2$-cells, with two $2$-cells being neighbors if they both attach to the same $3$-cell (see Fig.~\ref{fig:cube}).  In a dual picture, defect surfaces are now closed $d-2$-dimensional surfaces.

Ref.~\onlinecite{PercIsing} upper bounds $T_p$ by choosing a subset of Peierls contours which can be counted more easily.  These contours are obtained by constructing a sequence of $k$ spins starting from a given spin (say, at the origin of an infinite hypercubic lattice), flipping that spin, and then flipping each next spin in turn by shifting by distance $1$ in any of the positive coordinate directions.  For the given starting spin, there are $d^{k-1}$ such sequences of spins giving entropy $\sim k \log(d)$, while the area of the contour is proportional to $dk$.  So for $\beta \lesssim \log(d)/d$, the energetic cost does not suppress the appearance of these chains.
This bound generalizes straightforwardly to our problem, giving the same scaling $T_p \lesssim d/\log(d)$.  

\begin{figure}
\begin{center}
\scalebox{1}{\includegraphics[width=3.5in]{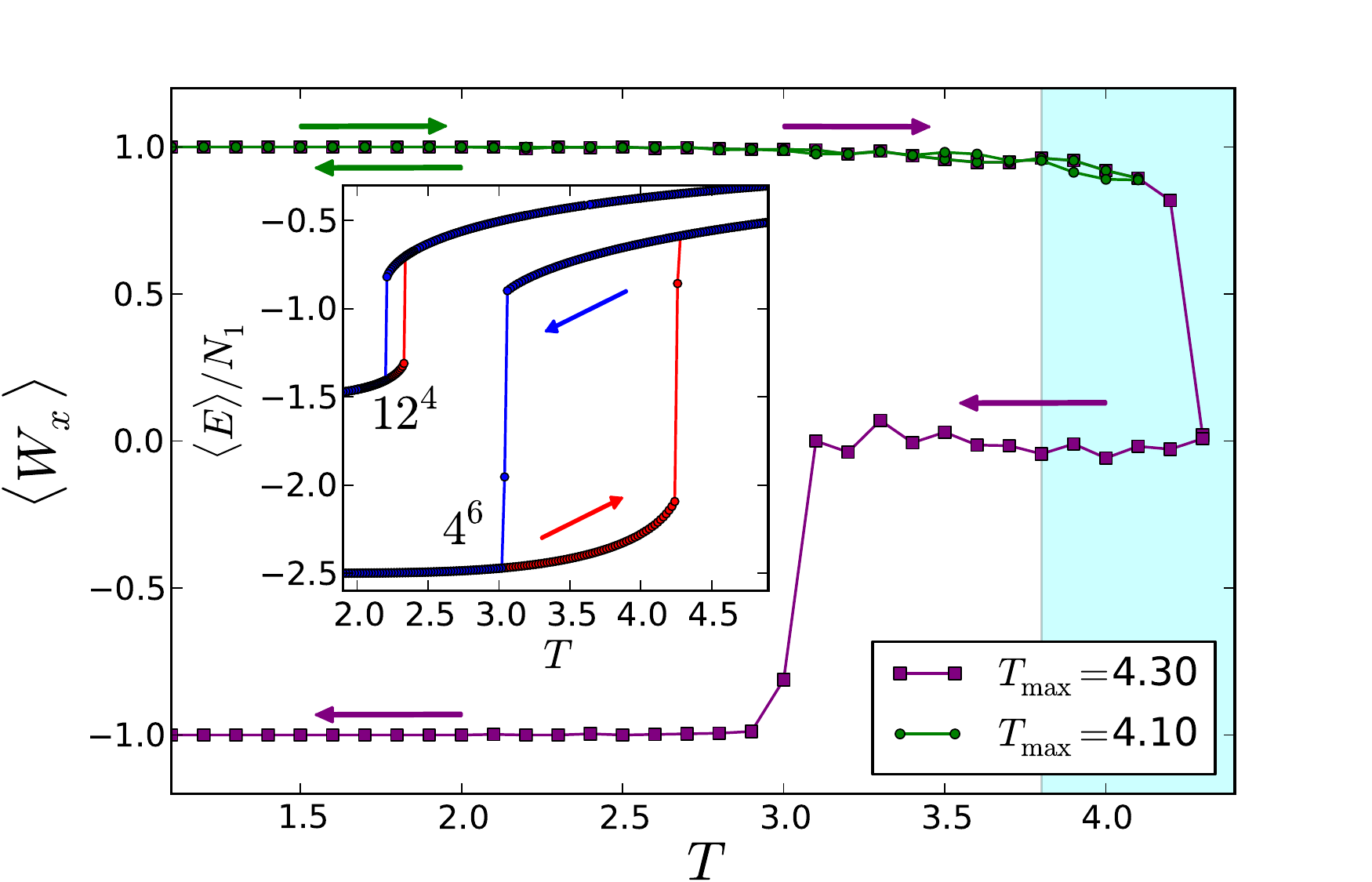}}
\end{center}
\caption{ 
Wilson loop in the $x$-direction for the $L=4$, $d=6$ code. 
Each line is a {\it single} Monte Carlo run,  where the simulation was warmed from $T=0$ to 
$T_{\rm max}$, then re-cooled to $T=0$.  Each data point represents the average over two thousand 
Monte Carlo steps at each temperature.  Averaging over many Monte Carlo runs would give a purple 
curve returning to $0$ at $T=0$.
%averaged over two thousand Monte Carlo steps at each temperature, 
%where the simulation was warmed from $T=0$ to  $T_{\rm max}$, the re-cooled to $T=0$.  
The region above $T_c=3.80$ is shaded.
Inset: internal energy of the $d=4$ and $d=6$ codes.  For $L=4$, $d=6$, the hysteresis jump occurs at $T=4.2$.
}
   \label{fig:energy}
 \end{figure}

We show below that for the $(1,d-1)$ code for large $d$,  $T_c \sim d$, so $T_c>T_p$ for sufficiently large $d$.  Thus, just as the Peierls argument cannot correctly predict $T_c$ in the Ising model, arguments based on percolating defects cannot correctly predict $T_c$ of certain high dimensional toric codes.
However, we claim that $T_c$, rather than $T_p$, determines the upper temperature at which the code is a self-correcting memory.  To act as a self-correcting memory, we need to define a recovery procedure. Then, we are interested in the question of encoding information into the memory at $T=0$, heating the memory to some $T<T_c$, allowing it to stay at that temperature for some time, and finally trying to recover the information (see Fig. \ref{fig:energy}).  In general, a recovery procedure involves measuring the set of defect plaquettes, called the ``syndrome".  Given the syndrome, additional spin flips are applied to correct the defects, obtaining a state without defects, and finally the encoded information is read.  Thus, from when the information is encoded to when it is read, some large number of spin flips occur as a combination of thermal noise and the recovery procedure.  Since these spin flips map from one ground state to another, they are a $1$-cocycle.  If the $1$-cocycle is topologically trivial, then the information can be recovered.  For a $(1,d-1)$ code, in a dual picture with spins on the $d-1$-cell, the spin flips are a cycle (a closed $d-1$-dimensional surface).  For $T<T_c$, the dynamics are not critical and the system relaxes quickly; thus, in the thermodynamic limit it is unlikely that the dynamics will create a topologically nontrivial $d-1$-dimensional surface of spin flips.

{\it Mean-Field Phase Diagram for $(p,q)$ Codes---}
We use mean-field theory for Hamiltonian $H_A$ to understand large $d$ behavior.  In contrast to Ref.~\onlinecite{MFTIsing}, no results will be proven on $T_c$, but the mean-field theory is still likely exact at large $d$.  From here on, since we consider a Hamiltonian which involves only $S^z$ operators, all our calculations are classical, considering only operators diagonal in the $S^z$ basis, and we set $J_A=1$.
The natural starting point for mean-field theory is a factorized probability distribution, $P(\{S^z_i\})=\prod_i p_i(S^z_i)$, giving the probability for a spin configuration as a product of the probabilities for each spin, as used in Ref.~\onlinecite{wipfbook} for the $Z_2$ gauge theory.

However, this method does not respect gauge invariance.
While this invariance only slightly changes $T_c$ for the $(1,d-1)$ code, it gives a large change in $T_c$ for $(d/2,d/2)$ codes where the gauge group is bigger.
Gauge invariance is the property that flipping all spins on $p$-cells attached to any given $p-1$-cell does not change the energy.  This leads to an extensive ground state entropy, while there is no way to obtain an extensive entropy in a zero temperature mean-field state.
By counting dimensions in a chain complex, the ground state degeneracy of $H_A$ can be estimated as
\be
\label{degen}
2^{N_{p-1}-N_{p-2}+N_{p-3}-...\pm N_0}
\ee
for a $(p,q)$ code,
up to $O(1)$ corrections which arise due to any nontrivial homology of the system. % This estimate arises by counting dimensions in a chain complex.\cite{Pauling}.

To make a gauge invariant mean-field theory we add constraints to $H_A$ to fix a unique representative from each orbit under the gauge group.  For a $(1,d-1)$ code, we use the gauge invariance to fix $S^z=+1$ for all $1$-cells oriented along some given lattice direction.  We then use a product ansatz for the remaining, unfixed spins.  We consider only infinite systems here; for a finite system with linear size $L$ with nontrivial homology we can fix all but a $1/L$ fraction of the spins.  Having fixed these, the remaining Hamiltonian has both two-spin and four-spin interactions.  The latter arise from plaquettes on which none of the $1$-cells are oriented in the $1$ direction, while the former arise from plaquettes with two $1$-cells in the $0$ direction and the other two not in that direction.
The mean-field equations are
\be
\label{MFT1}
\langle S^z \rangle=\tanh\Bigl(\beta(2\langle S^z \rangle + 2(d-2) \langle S^z \rangle^3)\Bigr),
\ee
where $\langle S^z \rangle$ is the average of $S^z$ on the unfixed $1$-cells.

The mean-field theory gives a variational lower bound for the free energy of the system with the spins fixed; subtracting $(1/\beta) N_0 \log(2)$ from this to account for gauge degeneracy gives a lower bound on the free energy $-(1/\beta) \log(Z_A)$.  Thus one could consider several different gauge fixings and choose the one that leads to the lowest free energy.  We do not do this here, but it will be useful for the mean-field theory for $(d/2,d/2)$ codes where the ground state degeneracy is much larger. For $(1,d-1)$ codes, the ratio between number of gauge group generators and number of spins goes to $0$ as $d \rightarrow \infty$, while for $(d/2,d/2)$ codes the ratio approaches $1/2$.

\begin{figure}
\begin{center}
\scalebox{1}{\includegraphics[width=3.5in]{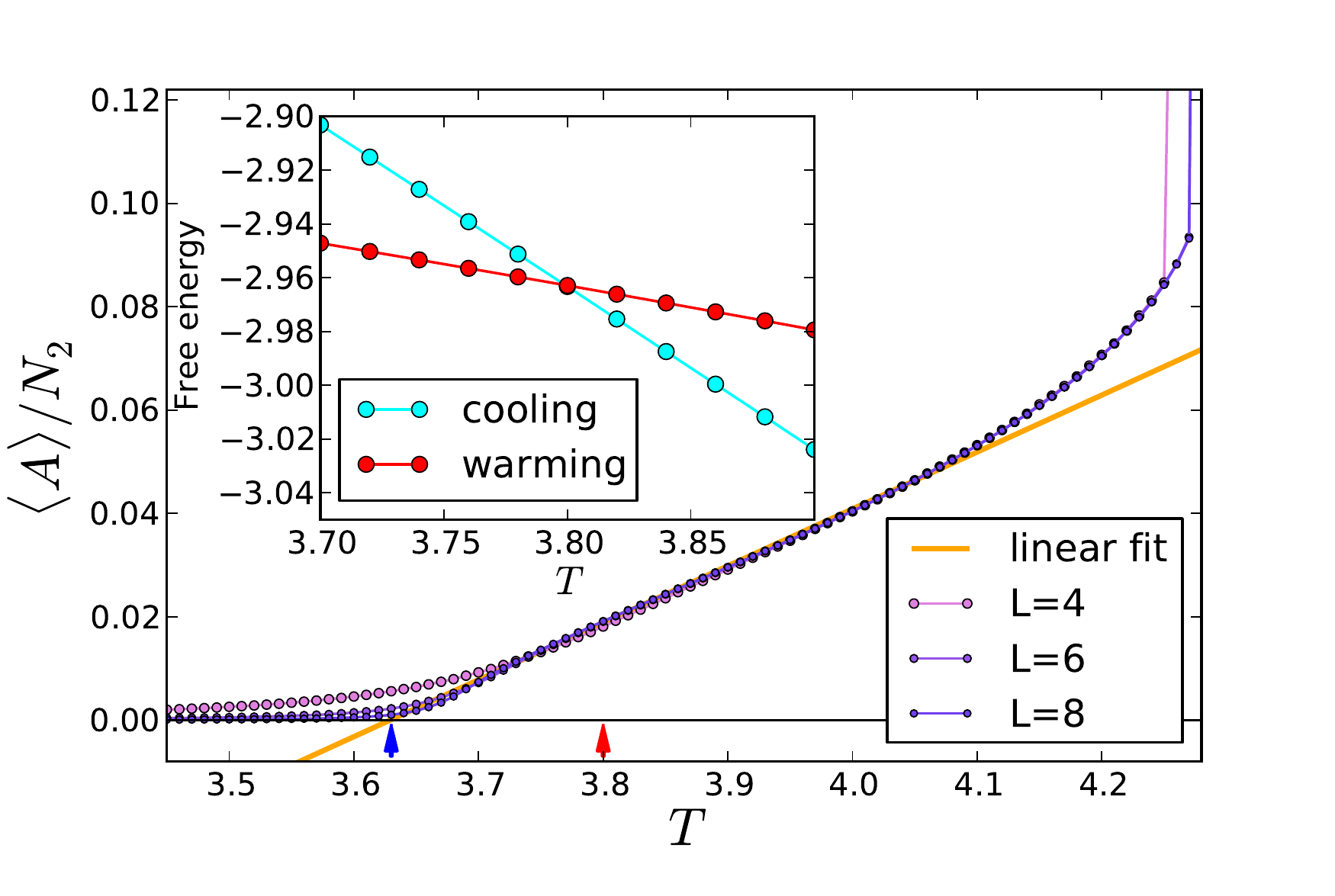}}
\end{center}
\caption{Onset of percolation in $d=6$.  From the crossing of a linear fit to the $L=8$ data with the axis, we estimate $T_p = 3.63$ (blue arrow on $T$-axis).  Inset: $T_c=3.80$ (red arrow on $T$-axis of main plot) from the crossing of the upper and lower branches of the free energy. }
   \label{fig:perc}
 \end{figure}

\begin{figure}
\begin{center}
\scalebox{1}{\includegraphics[width=2.9in]{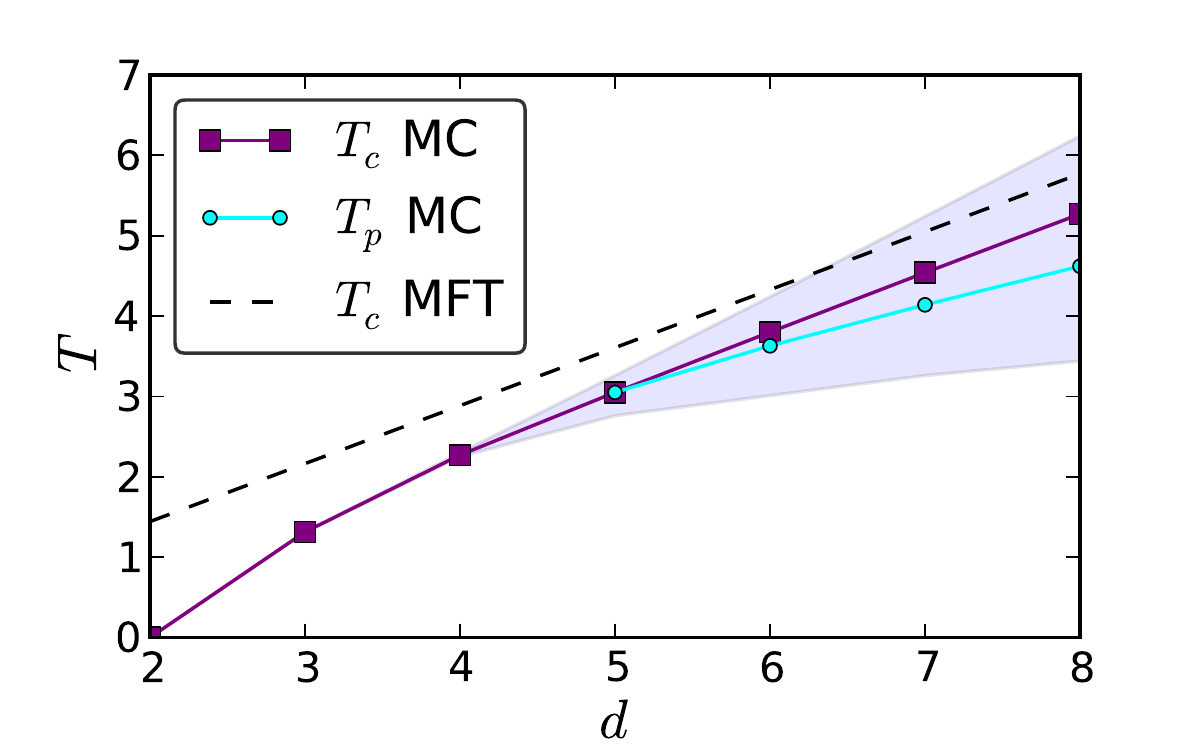}}
\end{center}
\caption{Transition temperatures versus dimension. The shaded region represents a typical region of hysteresis in $T_c$.  
%The dashed line is the equation for $T_c$ from mean-field theory. 
}
   \label{fig:Tc}
 \end{figure}

{\it Monte Carlo Measurement of $T_c$ and Hysteresis ---}
Using standard Metropolis Monte Carlo procedures, we simulate $H_A$ in Eq.~(\ref{Asum}) on 
various $L^d$ size lattices with periodic boundary conditions.
Standard thermodynamic estimators, e.g.~the specific heat, straightforwardly demonstrate that 
the $(1,2)$ code is a continuous critical point in the same universality class as the $3d$ Ising model \cite{KogutRMP}.
For $d \geq 4$, we observe a strongly first order phase transition, with hysteretic behavior as illustrated in Fig.~\ref{fig:energy}.
To obtain an accurate estimate of $T_c$, one must therefore measure the crossing of two free energy
branches, obtained by integrating the internal energy starting from high and low temperature (see inset of Fig.~\ref{fig:perc}).
This procedure is able to accurately reproduce $T_c$ for $3d$ and $4d$ (known from duality arguments \cite{wipfbook} to be 1.314 and 2.269 respectively) to the third decimal place with small system sizes.
Fig.~\ref{fig:Tc} shows the results for $T_c$ as a function of $d$, as well as a typical range of hysteresis
for $L=4$.

We illustrate how one can take advantage of the large hysteresis region to perform error correction even above $T_c$, by directly measuring the topological bit encoded in a Wilson loop, defined as $W_{\alpha} = \prod_i^L S^z_i$ for $i$ in a closed line in direction $\alpha$ (where $1 \le \alpha \le d$).  In Fig. \ref{fig:energy}, we plot $\langle W_x \rangle$ for two simulations -- one where the temperature is increased from $T=0$ to a temperature $T_{\rm max} > T_c$ but less than the hysteresis jump, the other for a $T_{\rm max}$ greater than the hysteresis jump.  Upon cooling from $T_{\rm max}$ to $T=0$, we observed a near 50 \% probability that the topological bit was destroyed when $T_{\rm max}$ was above the hysteresis jump (only one run is illustrated), while for the lower $T_{\max}$ the bit was always retained on the timescale of the simulation.

We estimate $T_c$ using mean-field theory.  The mean field equations always have a trivial solution with $\langle S^z \rangle=0$ and for sufficiently low $\beta$ there is also a nontrivial solution.  For large $d$, the nontrivial mean-field solution exists for $\beta/d\geq 2.017...$, as found numerically solving the equations.  The value of $\langle S^z \rangle$ at this $\beta$ is $0.889.$ We now consider the range of $\beta$ for which the nontrivial solution has lower free energy than the trivial solution.  The free energy of the trivial solution is $-T N_1 \log(2)$.  The free energy of the nontrivial solution requires a numerical calculation.  For an analytic estimate, we can use the free energy of the ground state sector, $-N_2-T N_0 \log(2)=-((d-1)/2-T\log(2)/d) N_1$, accounting for the gauge degeneracy, giving a crossing of the free energies at
\be
T_c=d/(2 \log(2)).
\ee
Corrections to this from a numerical solution of the equations are very small as $\langle S^z \rangle$ is close to $1$ at the given $T_c$.
The ratio $T_c/d$ in simulations is less than this value for a given $d$, but approaches it as $d$ increases
(see Fig.~\ref{fig:Tc}).

{\it Monte Carlo Measurement of Percolation---}
Monte Carlo simulations are able to give clear estimates of quantities relating to percolation, through measurement of the size of clusters of defect 2-cells (plaquettes with an odd number of $S_z = 1$ on the corresponding 1-cells).  As described above, any two 2-cells are defined as being neighbors if they share the same 3-cell (see Fig.~\ref{fig:cube}).
We can use this definition to measure quantities related to the size and topology of each defect cluster.  In order to identify the unique 
clusters in a simulation cell with a given number of defect 2-cells, we developed a variation of the standard Hoshen-Kopelman \cite{HK1}
algorithm for general higher-dimensional networks (see e.g.~Ref.~\cite{HK2}).  Using this, we measure the Monte Carlo average
of the {\it largest} cluster size, $\langle A \rangle$, as a function of temperature.

As illustrated in Fig.~\ref{fig:perc}, there is a strong linear onset of the largest cluster size at some temperature -- a clear sign
of percolation.  To take advantage of hysteresis of the low-temperature phase, simulations were started at $T=0$ and warmed 
until the discontinuity in the energy was observed.  The percolation transition $T_p$ is determined by a simple extrapolation
of the straightest region of the data (Fig.~\ref{fig:perc}).  The linear fit matches mean-field theory critical exponents for percolation as expected in $d \geq 6$.  For $d \ge 6$, we observe a clear separation of $T_p$ and $T_c$. 
For $d=5$, we also find a percolation transition in the metastable regime of the low temperature phase, 
however it appears that $T_p = T_c$ to within our simulation errors.  For $d=4$, no sign of percolation below $T_c$ was observed.

{\it Discussion}
We have studied various high dimensional toric codes, finding that percolation of thermally-activated defects $T_p$ occurs below the phase transition temperature $T_c$ in some cases.  
Contrary to the conventional understanding that 
percolating defects destroy the topological information at $T_p$,
we explicitly demonstrate that self-correction can occur for $T$ well above $T_p$, where percolating defects exist.

Further, Monte Carlo simulations of thermal spin flip processes discover a very large region of first-order hysteresis near $T_c$.
That is, the low temperature phase is stable on numerically accessible time scales well above the true $T_c$.
This allows for the possibility that the self-correcting phase may survive for $T$ well above $T_c$, in the ``superheated'' regime.
Of course, for $T>T_c$, eventually a bubble of the high temperature phase must nucleate within the low temperature phase, leading to equilibration, but the time scale associated with the formation of this bubble can be exponentially long.  
Thus, possibly a combination of active correction and self-correction could maintain a stable quantum memory well above $T_c$.  

Conversely, in physical implementations of surface codes, for example using superconducting quantum circuits \cite{Mariantoni},
syndrome qubits must be repeatedly read in order to actively correct errors that occur in data qubits encoding the topological
information \cite{Fowler1,Poulin}.   
It is therefore possible that the right combination of circuit size and dimension could lead to a wide temperature regime 
where computational effort for the classical processing required in this active error correction is significantly reduced.

{\it Acknowledgments:}
We thank P. Fendley, D. Poulin, and M. Mariantoni for valuable discussions, and especially L. Hayward and A. Kallin for collaboration on the Monte Carlo software.
This research was supported by NSERC of Canada and the Perimeter Institute for Theoretical Physics. 
%Research at Perimeter Institute is supported by the Government of Canada through Industry Canada and by the Province of Ontario through the Ministry of Research and Innovation.
We acknowledge the use of the computing facilities of the Shared Hierarchical Academic Research Computing Network (SHARC- NET:www.sharcnet.ca).


\begin{thebibliography}{99}
\bibitem{adjust} By adjusting the ratio $J_A/J_B$, one can obtain a $(2,d-2)$ code for which both phase transitions happen at the same temperature.
\end{thebibliography}
\end{document}